\documentstyle[prl,preprint,tighten,aps]{revtex}
\begin{document}
\draft
\title{Electron-phonon interactions on a single-branch quantum Hall edge}
\author{O. Heinonen$^{*,\dagger,}$\cite{UCF} and Sebastian Eggert$^*$}
\address{
$^*$Institute of Theoretical Physics and
$^{\dagger}$Department of Applied Physics\\
Chalmers University of Technology and G\"oteborg University,
S-412 96 G\"oteborg, Sweden}
\date{\today}
\maketitle
\begin{abstract}
We consider the effect of electron-phonon interactions on edge states
in quantum Hall systems with a single edge branch.
The presence of electron-phonon interactions
modifies the single-particle propagator for general quantum Hall edges,
and, in particular, destroys
the Fermi liquid even at integer filling. The effect of the
electron-phonon interactions may be detected
experimentally in the AC conductance or in the tunneling conductance between
integer quantum Hall edges.
\end{abstract}
A two-dimensional electron gas subjected to a strong perpendicular magnetic
field may exhibit the quantum Hall (QH) effect\cite{QHE_book}. The effect
occurs
because the electron gas is incompressible at certain densities,
which is due to an energy gap to bulk excitations.
In the integer QH effect
this gap is the kinetic energy gap (the cyclotron energy $\hbar\omega_c$)
in a magnetic
field, and in the fractional QH effect the gap arises because of the
electron-electron interactions. As a result of the bulk energy gap, gapless
excitations in the system can only exist at the edges. Such edge excitations
are density modulations localized at the edges of the system, and all the
low-energy properties of a QH system are determined by the edge excitations.

As was first demonstrated by Wen\cite{Wen1}, the density operators for
edge excitations obey a Kac-Moody algebra, similar in structure to that
obeyed by Luttinger liquids\cite{Luttinger1}. Because time-reversal
invariance is broken by the magnetic field, the excitations on one edge
can only propagate in one direction\cite{composite} corresponding to
`chiral
Luttinger liquids'\cite{Wen1}. Wen calculated the single-particle
propagator for the electron on the edge of a quantum Hall system at
filling factor $\nu=1/(2m+1)$ and showed that it is given by\cite{Wen1}
$G(x,t)\propto(x-v_Ft)^{1/\nu}$. Here, $x$ is a coordinate
along the edge, and
$v_F$ is the edge velocity determined by the details of the
potential confining the electron gas at the edge.
It is worth noting the contrast between chiral Luttinger liquids and
regular Luttinger liquids in one-dimensional (1D) interacting electron
systems without magnetic fields. In the regular Luttinger liquid,
the Fermi surface
is destroyed by repeated backscattering, and the exponent of the
single-particle propagator depends on the details of the electron-electron
interactions.
On the other hand, a chiral Luttinger liquid at a QH edge is
caused by the strong
correlations in the bulk, and the exponent of the
single-particle propagator is fixed by
the topological order in the bulk of the system.

The appearance of the anomalous exponent in the single-particle
propagator has important experimental implications.
It was shown by Wen\cite{Wen2} and
by Kane and Fisher\cite{KF1} that the tunneling conductance between
two edges of an FQHE system at bulk filling factor $\nu=1/(2m+1)$
depends on temperature as $T^{2(1/\nu-1)}$. The resonant tunneling
conductance was calculated by Moon {\em et al.}\cite{Moon1},
and Fendley {\em et al.}\cite{Fendley}, and measured by Milliken
{\em et al.}\cite{Milliken}. The experimentally measured tunneling
conductance does indeed exhibit a $T^{2(1/\nu-1)}$-dependence,
except for at the very lowest temperatures, where Coulomb interactions
between the edges may modify the conductance\cite{Moon2}.

In this letter,
we consider the effect of electron-phonon interactions on the QH edge states
of spin-polarized QH systems with bulk filling factor $\nu=1/m$, with $m$
an odd integer. Such systems have a single branch of edge excitations on each
edge.
Electron-phonon interactions in regular Luttinger liquids
have been considered previously\cite{Voit,Martin}.
Martin and Loss\cite{Martin} showed that coupling the electron
system to acoustic phonons destroys the Fermi surface, even in the absence
of electron-electron interactions. This effect is only appreciable if
the Fermi velocity $v_F$ is of the same order as the sound velocity $v_s$,
which may be achieved in strongly correlated electron systems where the
role of the Fermi velocity $v_F$
is played by the charge velocity of the Luttinger liquid.
The effect may also be appreciable in QH systems, where $v_F$ is determined
by the stiffness of the confining potential and the electron density, both
of which may be tuned electrostatically by gates.
We will show here that the electron-phonon
interactions will modify the single-particle propagator even in a
{\it chiral} Luttinger Liquid.   In particular, the electron propagator of
QH edge state will be shown to have the form
\begin{equation}
G(x,t)\propto
{1\over (x- v_\alpha t)^{T_{11}^2/\nu}}
{1\over (x-v_\beta t)^{T_{12}^2/\nu}}
{1\over (x+v_\gamma t)^{T_{13}^2/\nu}},
\label{prop2}
\end{equation}
where $v_\alpha$ is a renormalized edge velocity, and $v_{\beta,\gamma}$
are renormalized sound velocities. As a consequence,
even the integer quantum Hall edge states will not be Fermi liquids in the
presence of electron-phonon interactions.

The modification of the single-particle propagator by the
electron-phonon interactions may be detected experimentally, and we will
discuss two possibilities.
First, the AC conductance will
have resonances at (longitudinal) wave-vectors $q$ and frequencies $\omega$
related by $q=\omega/v_\alpha, \ q=\omega/v_\beta,$ and
$q=-\omega/v_\gamma$ which may in principle be
resolved and detected in an experiment. On the other hand, the
DC Hall conductance is {\em not} modified by the electron-phonon interactions.
Second, the anomalous exponents in the single-particle propagator will
modify the tunneling conductance and can in principle be measured, for
example in tunneling between two $\nu=1$ edges in a bilayer
system\cite{bilayers} with
an overall filling of $\nu_{\rm tot}=2$.
In the absence of electron-phonon interactions the edges of such a system are
 (chiral) Fermi liquids which
propagate in the same direction. Coulomb interactions alone cannot change
the temperature dependence of the tunneling conductance from $T^0$. Therefore,
{\em any} temperature behavior of the tunneling conductance at
sufficiently low temperatures must be due to
the electron-phonon interactions.
The electron-phonon interaction also modifies the temperature dependence
of the tunneling conductance between counterpropagating edge
states at very low temperatures
which has been measured by Milliken {\em et al.}\cite{Milliken}

The edge excitations in a quantum Hall system with filling fraction
$\nu=1/(2m+1)$ can be described by density modulations of an effectively
one-dimensional system\cite{Wen1}
\begin{equation}
H_e \equiv \frac{2 \pi}{L}\frac{v_F}{\nu} \sum_{k>0}
\ J_k J_{-k}, \label{H_e}
\end{equation}
where the densities $J_k$ obey commutation relations depending on
the filling
fraction $\nu$: $[J_k, J_{k'}] = - \frac{L\nu k}{2\pi} \delta_{k,-k'}$.
Here, $k$  is the wave-vector along the edge.
This theory represents a single $U(1)$ Kac-Moody algebra and
it is well known how to ``bosonize'' it in terms
of a chiral boson $\phi_R(x)$.
The current density $j(x)$ is written as
\begin{equation}
j(x) = \sqrt{\frac{\nu}{\pi}} \
\frac{\partial \phi_R}{\partial x}. \label{bos2}
\end{equation}
Using $J_k = \int dx \ e^{ikx} j(x)$, we can immediately express the
Hamiltonian in terms of the annihilation and creation operators
of the boson mode expansion:
\begin{equation}
a^\dagger_k = \sqrt{\frac{2 \pi}{L\nu k}} J_k,
\ \ \ a_k = \sqrt{\frac{2 \pi}{L\nu k}} J_{-k}
\end{equation} (the zero modes are omitted).
The full (spin-polarized) electron field can also be written in terms
of the chiral boson and has been identified as\cite{Wen1}
\begin{equation}
\psi(x) \propto \exp\left(i \sqrt{1/4 \nu \pi}\phi_R\right)\label{bos1}
\end{equation}
(this is not to be confused with the quasi-particle field $\chi$
which can also be defined in terms of the chiral boson
$\chi \propto e^{i \sqrt{\nu/4 \pi} \phi_R}$, but carries fractional charge).

We now consider the interaction of such a system
with phonons
\begin{equation}
H_{ph} = v_s \sum_{\vec{k}}
 \ |\vec{k}| b_{\vec{k}}^\dagger b_{\vec{k}}^{}. \label{H_ph}
\end{equation}
One normal mode of the phonons is assumed to be along
the quantum Hall edge, so that the crystal displacement $d(x)$
 along this edge can be expressed as
\begin{equation}
d(x) = \sum_k i (2 L \rho v_s k)^{-1/2} e^{i k x} \ (b_k + b_{-k}^\dagger).
\label{d_x} \end{equation}
where $\rho$ is the linear mass density of the crystal.
The electron-phonon interaction then becomes
\begin{equation}
H_{e-p} = D \int dx \rho(x) \partial_x d(x) =
v_c \int dk \ k (a_k^{} b_k^\dagger + a_k^\dagger b^{}_k), \label{H_e-p}
\end{equation}
where $D$ is the deformation potential constant and the coupling
$v_c = D\sqrt{{\nu/\pi \rho v_s}}$ is independent of $k$.
For specific values of these parameters, we consider a QH edge in a
GaAs heterojunction. We assume that the edge is along one of the
cubic axes of GaAs so that the piezoelectric coupling vanishes and can
be ignored. For electrostatic confinement by an electrode
with potential $V_g$, $v_F$ is of the
order of $\omega_c\ell_B^2/\ell$, where
$\ell_B=\left[\hbar c/(eB)\right]^2$ is the magnetic length and
$\ell=V_g\epsilon/(4\pi^2 n_0 e)$
is the length scale of the electrostatic confining
potential\cite{Chklovskii}
($\epsilon$ is the static dielectric constant
and $n_0$ is the two-dimensional electron density).
For a magnetic field strength of about 5 T and a density of
$n_0=10^{15}$ m$^{-2}$, this gives a Fermi velocity approximately
equal to the average sound velocity $v_s\approx5\times10^3\rm m/s$ in
GaAs. Thus, it should be possible to optimize the effects of electron-phonon
interactions in GaAs heterojunctions under ordinary conditions.
The deformation potential constant $D$ is approximately 7.4 eV
\cite{Levinson}. Assuming an effective cross-sectional area of $10^{-14}$
m$^{-2}$ of the
GaAs phonon system in the direction perpendicular to the electron propagation,
we then arrive at a coupling velocity $v_c/v_s\sim 0.1$.

\begin{figure}
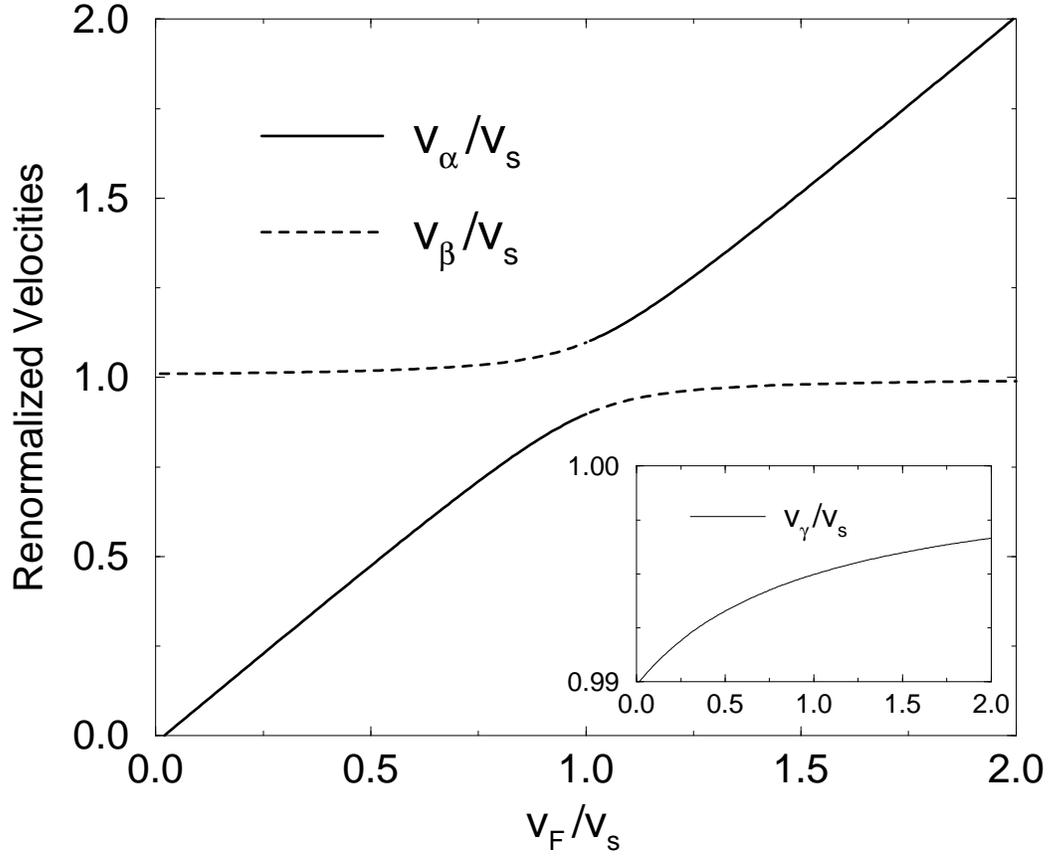

\caption{The renormalized velocities $v_\alpha$ and $v_\beta$ as
a function of the Fermi velocity $v_F$. The sound velocity $v_s$
sets the overall scale
and the coupling has been chosen to be $v_c/v_s = 0.1$.  The inset shows
the renormalized velocity of the counterpropagating mode. } \label{v}
\end{figure}

At this point it is straightforward to diagonalize the complete Hamiltonian
\begin{equation}
H = \sum_{k>0} k \left( v_F a^\dagger_k a_k^{} +
v_s[b^\dagger_k b^{}_k + b^\dagger_{-k} b^{}_{-k}]
 +  v_c [ a^\dagger_k (b_k+b^\dagger_{-k}) + h.c.]\right) \label{H}
\end{equation}
by using a generalized Bogliubov transformation $T$
\begin{equation}
(a_k, b_k, b^\dagger_{-k}) =  T\cdot \left(\begin{array}{c}
\alpha_k \\ \beta_k \\ \gamma^\dagger_{-k}\\ \end{array}\right)
\label{bogliubov}
\end{equation}
where $T$ is given in terms of three variables $\phi, \theta, \eta$:
\begin{equation}
T = \left(\begin{array}{ccc}
\cos\phi \cosh\theta & \sin\phi \cosh\theta\cosh\eta + \sinh\theta\sinh\eta
& -\sin\phi \cosh\theta \sinh\eta -\sinh\theta \cosh\eta \\
-\sin\phi & \cos\phi\cosh\eta & -\cos\phi  \sinh\eta  \\
-\cos\phi \sinh\theta & -\sin\phi \sinh\theta\cosh\eta - \cosh\theta\sinh\eta
& \sin\phi \sinh\theta \sinh\eta -\cosh\theta \cosh\eta \\
\end{array}\right).
\end{equation} 
The  Hamiltonian is now written as
\begin{equation}
H = \sum_{k>0} k  \left(
\alpha_k , \beta_k , \gamma^\dagger_{-k}\right)\cdot A
\cdot \left(\begin{array}{c}
\alpha_k^\dagger \\ \beta_k^\dagger \\ \gamma^{}_{-k}\\ \end{array}\right)
\label{rotated_H}
\end{equation}
where the coupling matrix $A$ is given by
\begin{equation}
A\equiv  T^\dagger\cdot \left(\begin{array}{ccc} v_F & v_c & v_c \\
v_c & v_s & 0 \\ v_c & 0 & v_s \\  \end{array} \right)\cdot T
\end{equation}
For the Hamiltonian to become diagonal, the off-diagonal elements of $A$
are required to vanish, which determines the three angles
$\phi, \theta, \eta$ and in turn also the diagonal elements of $A$ (i.e.
the renormalized Fermi and sound velocities). Moreover, the boson
representing the electron density now becomes according to
Eq.~(\ref{bogliubov})
\begin{equation}
a_k = T_{11} \alpha_k + T_{12} \beta_k + T_{13} \gamma_{-k}^\dagger
\label{rescale}
\end{equation}
Therefore, the electron field in Eq.~(\ref{bos1}) must now be
expressed in terms of three independent boson fields, namely a right
moving ``charge'' boson $\alpha_k$ and one right and one left moving
``sound'' boson $\beta_k, \ \gamma_k$.
The electron propagator then becomes a product of
three factors according to Eqs.~(\ref{bos1}) and (\ref{rescale})
\begin{equation}
G(x,t)   \equiv   <\psi^\dagger(x,t)\psi(0,0)>
 \propto  {1\over (x- v_\alpha t)^{T_{11}^2/\nu}}
{1\over (x-v_\beta t)^{T_{12}^2/\nu}}
{1\over (x+v_\gamma t)^{T_{13}^2/\nu}},
\label{propagator}
\end{equation}
where $v_\alpha=A_{11}$ is the renormalized Fermi velocity and
$v_\beta =A_{22}, \ v_\gamma = A_{33}$ are the renormalized sound
velocities.  In Fig.~\ref{v} we have plotted the renormalized velocities
$v_i \ vs. \ v_F \  (i=\alpha,\beta,\gamma)$
for a coupling of $v_c/v_s=0.1$.  The total `momentum' $v_\alpha
+v_\beta-v_\gamma = v_F$ of the electron is conserved by the
transformation $T$.  The right propagating velocities $v_\alpha$ and
$v_\beta$ show a discontinuity at the resonance $v_F=v_s$ of equal magnitude
which appears to be quadratic in the coupling $v_c/v_s$.  The left
propagating velocity $v_\gamma$ is only slightly modified of the order
of 1\% from its original value $v_s$.

\begin{figure}
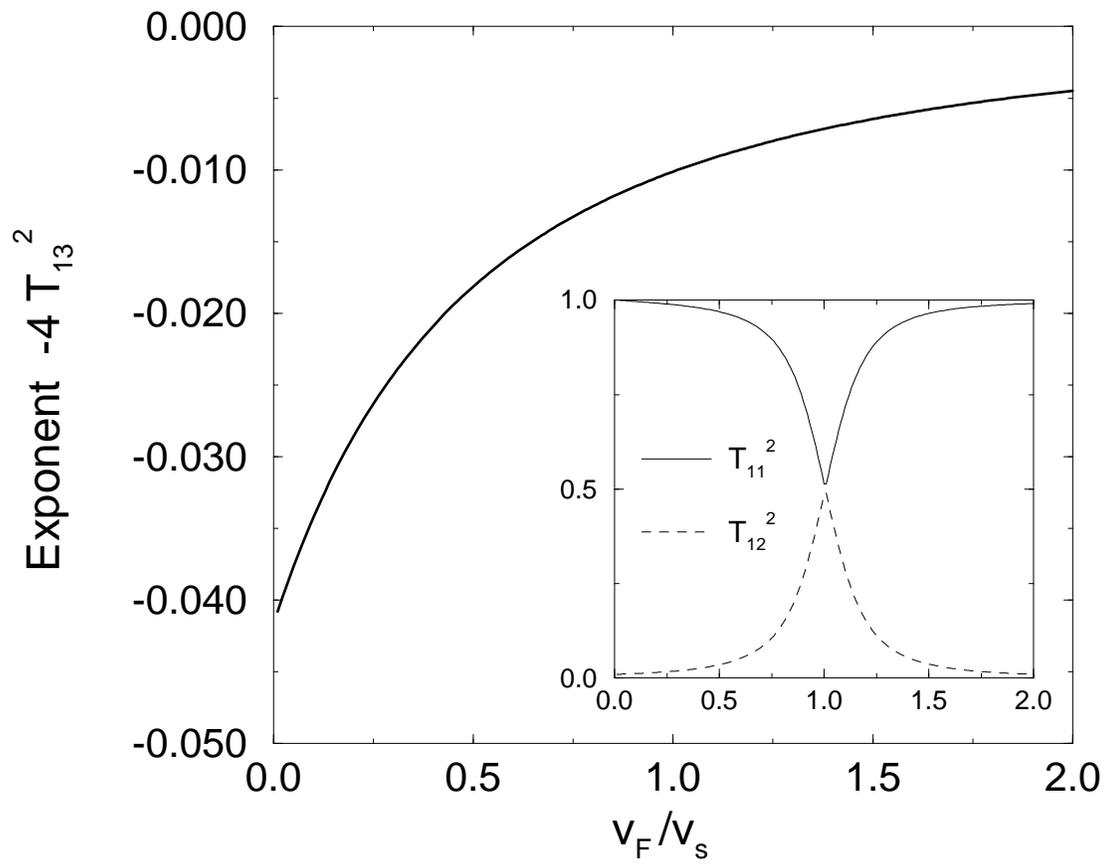

\caption{The exponent of the temperature dependence of the tunneling
conductance as a function of the Fermi velocity $v_F/v_s$ for a
coupling of $v_c/v_s = 0.1$.  The inset shows the `spectral weights' of
the forward propagating modes.} \label{T}
\end{figure}

The fact that the electron propagator breaks up into three pieces,
corresponding to the normal modes of the Hamiltonian Eq.~(\ref{rotated_H}),
will have experimental consequences for transport properties. We first
calculate the linear response to a scalar potential
$\phi(x,y)=-Ey\cos(qx-\omega t),$ where we take $q,\omega>0$. This
potential gives an electric field ${\bf E}(x,y)=
E\left[-qy\sin(qx-\omega t)\hat{\bf x}+\cos(qx-\omega t)\hat{\bf y}\right],$
where $y$ can be taken to be constant.
The perturbed charge density in response to the potential is then
obtained as
\begin{equation}
\delta\rho(x,t)  =
\frac{e^2Ey}{h}\nu q \cos(qx-\omega t)
 \left[{T_{11}^2\over(v_\alpha q-\omega)}
+{T_{12}^2\over (v_\beta q-\omega)}+{T_{13}^2\over(v_\gamma q+\omega)}\right].
\end{equation}
By using the continuity equation $\partial\rho/(\partial t)+
\partial j(x)/(\partial x)=0$, we can then obtain the current response function
to the applied potential as
\begin{equation}
\widetilde\sigma(q,\omega)=\frac{e^2\nu}{h}\omega
\left[{T_{11}^2\over(v_\alpha q-\omega)}+{T_{12}^2\over(v_\beta q-\omega)}
+{T_{13}^2\over(v_\gamma q+\omega)}\right].
\label{sigma}
\end{equation}
Experiments dictate that the DC Hall conductance
$\sigma_H=\lim_{\omega\to0}\lim_{q\to0}\widetilde\sigma(q,\omega)$
must not be altered from its quantized value $e^2\nu/h$, which is indeed the
case according to Eq.~(\ref{sigma}) since the matrix elements obey
the sum rule $T_{11}^2 + T_{12}^2 - T_{13}^2 = 1$.
On the other hand, the AC conductance will
exhibit resonance structures when
$\omega=v_\alpha q, \  \omega=v_\beta q,\ $ and $\omega=-v_\gamma q$
in response to a potential $\phi(x,y)$.
Provided at least two of the `spectral
weights' $T_{11}^2$, $T_{12}^2$, and $T_{13}^3$ are not too small, and
the corresponding renormalized velocities  are
not too close, these resonances can then in principle be resolved and
detected. In Fig.~\ref{T}  we see that near
$v_F/v_s \sim 1$, both $T_{11}^2$ and $T_{12}^2$ are close to 0.5, while
$v_\alpha$ and $v_\beta$ are on opposite sides of $v_s$ (Fig.~\ref{v}).
With an experimentally
reasonable value of $q\sim10^5$ m$^{-1}$ and
$v_s\sim v_F \sim10^3\rm m/s$, this gives a resonance at
about $10^8$ Hz, well within experimentally accessible range.
Figure \ref{T} also shows that the `total spectral weight' of the
electron consists mostly of the forward propagating modes,
which contribute almost equally at resonance $v_F = v_s$.  This means
that only very little charge is transported in the counterpropagating
 direction which was to be expected.

Since the single-particle propagator is changed by the electron-phonon
interaction according to Eq.~(\ref{propagator}), the
single-particle density-of-states and properties depending on it
such as the tunneling conductance will also be affected by the
coupling to the phonons.
In particular, we consider the inter-layer tunneling conductance between
two edges of a bilayer system with integer filling in each layer, and a
total filling factor of $\nu_{\rm tot}=2$. This can in principle be measured by
attaching probes to the different layers separately in a system
with large enough separation between the layers that the bulk
tunneling probability vanishes. A gate at the edge can then be used
to adjust the tunneling probability between the edges.
At low enough voltage across the tunneling junction, tunneling through the
bulk will be suppressed, and only the edge tunneling current appreciable.
Note that in this arrangement, the two edges propagate in the same direction.

The tunneling current is determined by the retarded response
function\cite{Wen2}
\begin{equation}
X_{\rm ret}(t)=-i\theta(t)\langle\left[
A(t),{A}^\dagger(0)\right]\rangle,
\label{Xret}
\end{equation}
where $\Gamma  A=\Gamma\psi_1(x=0){\psi_2}^\dagger(x=0) + h.c.$
is the tunneling operator, with $\Gamma$ the tunneling amplitude, and
$\psi_i(x)$, $i=1,2$, the electron field operator on the two edges.
{}From Eq.~(\ref{propagator}) and following Wen\cite{Wen2}
it is a straightforward exercise to determine
the temperature dependence of the tunneling differential
conductance, with the result that
\begin{equation}
\left.{dI_t\over dV_t}\right|_{V_t=0}\ \ \propto \ \
T^{2(T_{11}^2+T_{12}^2+T_{13}^2)-2} \ \ = \ \ T^{-4 T_{13}^2}.
\label{tunnel}
\end{equation}
Due to the fact that $T_{13}^2\not=0$ in the presence of electron-phonon
interactions, the differential tunneling conductance
will now depend on temperature, in fundamental contrast to the tunneling
between two chiral Fermi liquids propagating in the same direction, which
is temperature {\em independent} at low temperatures, even in the presence
of electron-electron interactions between the two edges.
{}From Fig.~\ref{T} we see that the magnitude of the exponent $-4 T_{13}^2$
for the parameters chosen here is of the order of $10^{-2}$, which is
small (the exponent appears to be quadratic in the coupling constant
$v_c/v_s$).  However, the main point is that any measured
temperature dependence at all will be due to the
electron-phonon interactions.

In conclusion we have shown that the electron-phonon interaction
modifies the chiral Luttinger Liquid on the quantum Hall edge.
The single particle propagator becomes a product of three separate
modes, one of which is  always counterpropagating.  From this we have
predicted direct experimental consequences for the AC conductivity
of the quantum Hall bar
and the temperature dependence of the tunneling conductance.

\begin{acknowledgements}
The authors would like to thank Henrik Johannesson and Micheal Johnson
 for helpful discussions.
 O.H. expresses his gratitude to Stellan \"Ostlund and
Mats Jonson at Chalmers University of Technology and G\"oteborg
University for their hospitality and support during a great
sabbatical stay, and
to the National Science Foundation for support through grant DMR93-01433.
This research was supported in part
by the Swedish Natural Science Research Council.
\end{acknowledgements}


\begin{references}
\bibitem{composite} We are here not considering systems with composite edges,
which may have counterpropagating modes.
\end{references}
\end{document}